\documentclass[aps,prl,twocolumn,groupedaddress,floatfix]{revtex4}

\usepackage[dvips]{graphicx}
\begin{document}

%\preprint{}

\title{Better Lunar Ranges with Fewer Photons - Resolving the Lunar Retro-reflectors}


\author{Louis K. Scheffer}
\email[]{lou@cadence.com}
%\thanks{}
\affiliation{Cadence Design Systems}


\date{\today}

\begin{abstract}
Because of lunar librations, the retroreflectors left on the moon do not, 
in general, face directly at the Earth.  Usually this is regarded as a 
disadvantage.  It results in a spread of arrival times,  because each
cube that comprises the retroreflector is at a slightly different distance
from the earth.  However, we can turn this same effect into an advantage. 
Using pulses and detectors somewhat faster than those currently used for lunar ranging, 
we can resolve at least some of the structure of a retroreflector, at least when
the libration angles are large.  This additional
structure in the transfer function means that
a unique mm level fit can be obtained with many fewer photons.  
Fitting to the expected reflector
transfer function in general requires fewer photons than straight averaging, and smoothly
reduces to averaging in the cases where no structure can be resolved.  
The gains from resolving the reflectors are largest at
large libration angles, exactly the case where averaging is most inefficient.  
In these cases the number of photons needed can be reduced by an order of magnitude or more.
Analysis shows that angles for which the gain is very high happen several times 
each month, with the details depending on the exact librations.  
Experimental validation
of this technique should be possible with existing SLR stations and mockups of the
lunar reflectors.
\end{abstract}

%\keywords{}

\maketitle

\section{Introduction}
Although the moon is tidally locked to the Earth, the Earth is not always in the exact same position in the lunar sky.  
Because the moon's orbit is eccentric, the Earth appears to move about 7.7 degrees east and west around its mean position over the course of an orbit.  
Because the moon's orbit is tilted with respect to the ecliptic, the Earth 
appears to move about 5.5 degrees north and south of its nominal 
position as well.  These effects together are known as lunar librations.

These lunar motions have practical implications for Lunar Laser Ranging (LLR).  
The reflectors left on the moon are are fixed with respect to the surface and do not 
track the earth.  
If the reflectors on the moon were comprised of a single corner cube, this misalignment would
matter very little, since corner reflectors by design have equal path lengths to
any point on their surface, for any angle.  However, in order
to get enough cross section for a practical signal strength, while keeping the
weight low, the lunar retroreflectors consist of a planar array of corner cubes.  
When the normal to this plane does not pass through the ranging telescope 
(which it normally does not), 
this adds dispersion to the returned pulse, 
since the individual corner cubes are at different distances from the 
telescope.  
\section{The lunar retroreflectors}
There are 4 commonly used retro-reflectors on the moon, of three different designs.
Since the details are crucial to our discussion, the three designs are reviewed here.
\subsection{The Apollo 11 and 14 reflectors}
\begin{figure}
\begin{center}
\noindent
\includegraphics[height=7.5cm]{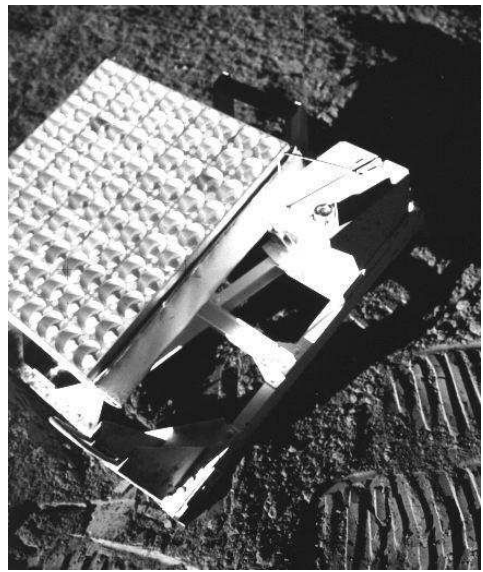}
\caption{The Apollo 14 retro-reflector on the moon.}
\label{fig:A14}
\end{center}
\end{figure}

The Apollo 11 (A11) and Apollo 14 (A14) reflectors are square, with 10 rows and 10 columns of circular 
corner cubes, each 38 mm in diameter.  
The spacing from row to row and column to column is roughly 46 mm.
A photo of the A14 reflector on the moon is shown in Fig. \ref{fig:A14}.
\subsection{The Lunakhod reflectors}
\begin{figure}
\begin{center}
\noindent
\includegraphics[height=4cm]{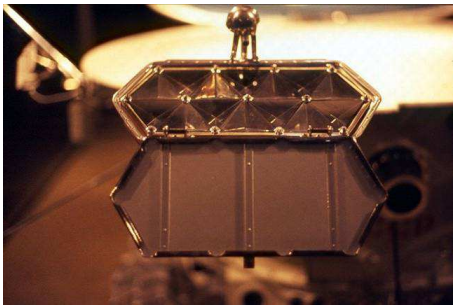}
%\caption{Line drawing of Lunakhod reflector.  The triangles are the corner cubes; their optical centers are marked with crosses.  Each triangular corner cube is 106 mm on a sde.}
\caption{Photo of the Lunakhod reflector.  The triangles are the corner cubes. Each triangular corner cube is 106 mm on a side.  From \protect{\cite{Lunokhod1}} }
\label{fig:Lunakhod}
\end{center}
\end{figure}

These reflectors feature 14 triangular corner cubes, arranged in a hexagonal packing.  
Each triangle is about 106 mm on a side.  The hexagonal design yields 6 fold symmetry rather than the 4 fold of the A11 design.
Since the Lunakhod reflector is not as well aligned to earth (it is believed to be
about 5 degrees off from the mean Earth position\cite{OCA}), there will be at least 
some times when the deviation from normal is higher than it will ever 
be for a better aligned array, such as those of Apollo 
(believed to be within about 1 degree of the mean position\cite{OCA}).
A photo of a Lunakhod retroreflector is shown in Fig. \ref{fig:Lunakhod}.
\subsection{The Apollo 15 reflector}
\begin{figure*}
\begin{center}
\noindent
\includegraphics[height=7.5cm]{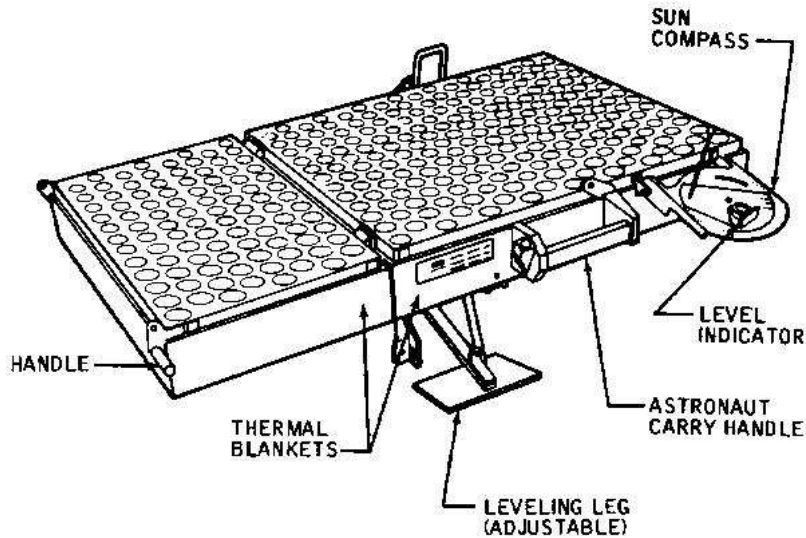}
\caption{Line drawing of A15 reflector, from \protect{\cite{MySpace}}.  Note the gap between the two sections.}
\label{fig:A15Line}
\end{center}
\end{figure*}

\begin{figure}
\begin{center}
\noindent
\includegraphics[height=7.5cm]{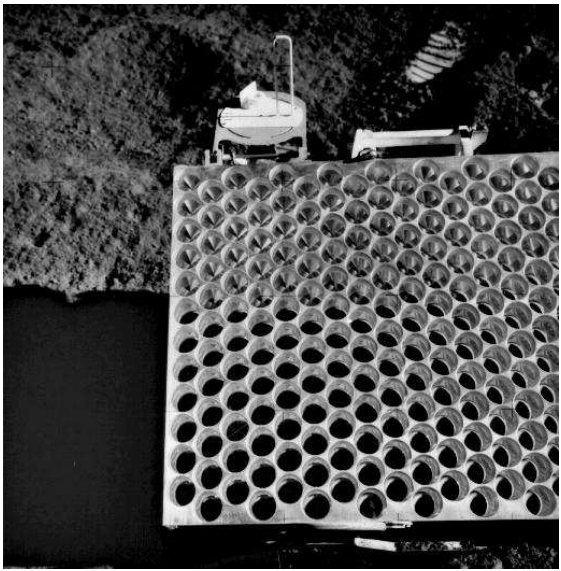}
\caption{Photo of the A15 reflector as left on the moon.}
\label{fig:A15Photo}
\end{center}
\end{figure}

This is a larger reflector, with 300 corner cubes in two panels, using close 
(hexagonal) packing.  
Each individual reflector is similar to those of Apollo 11 and 14 - 38 mm in diameter.
One panel has 204 reflectors in 17 columns of 12 rows, then there is a gap, then there is another panel of 96 reflectors in 8 columns of 12 rows.  The total array dimensions are 1.05 meters by 0.64 meters.  
Fig. \ref{fig:A15Line} is a drawing of the whole apparatus.
Fig. \ref{fig:A15Photo} is a photograph of the reflector on the moon, showing the hexagonal close packing.

\section{The problem}
To understand the
order of magnitude of the problem, a meter sized reflector with a 7.7 degree 
misalignment has path lengths that vary by up to
26 cm depending on which cube reflects a given photon.
This results in about a $\pm 430$ ps spread in arrival times.  
A more detailed analysis, assuming each photon has an equal chance of being
reflected by each corner cube, and using the physical dimensions of the A15 array, is shown in Table 1.
This table shows the dispersion of the Apollo 15 array as a function of the east-west and north-south librations.
This is consistent with the conclusions of the OCA group\cite{OCA}. (In this plot, and all tables and plots to follow, only the first quadrant of possible librations is shown. 
This is sufficient since for all of the reflectors considered here, the other quadrants
are just mirror images of the first quadrant, reflected about the X or Y axis, or both.)
\begin{table}
\begin{center}
\begin{tabular}{|c||c|c|c|c|c|} \hline
 8.0&  148.0&  163.3&  202.4&  254.5&  312.9\\ \hline
 6.0&  111.1&  131.2&  178.2&  236.6&  299.5\\ \hline
 4.0&   74.2&  102.1&  158.6&  222.8&  289.4\\ \hline
 2.0&   37.1&   79.6&  145.5&  214.1&  283.2\\ \hline
 0.0&    0.0&   70.5&  140.9&  211.1&  281.1\\ \hline \hline
 & 0.0& 2.0& 4.0& 6.0& 8.0\\ \hline
\end{tabular}
\label{tab:ps}
\end{center}
\caption{Dispersion in ps (rms) as a function of the east-west and north-south angles (in degrees) for the Apollo 15 array.  The same angle North-South causes less dispersion since the array is wider than it is tall.}
\end{table}

With modern ranging systems this dispersion is the largest component of 
ranging error budget, as the other errors have been reduced dramatically
since the early days of Lunar Laser Ranging.
The total of all other uncertainties is only 60-75 ps for the OCA station, 
and is estimated as 52 ps for the Apache Point effort \cite{Murphy1}
(currently under construction).  
The OCA paper \cite{OCA}, Figure 7, has a very nice plot showing that the received
dispersion is in fact very close to what would be predicted from the 
reflector mis-alignment at the time of measurement.  They conclude that
the combined effect of all system jitter, excluding only the
retroreflector, is only 60 ps.
``Numerically, one gets: $\sigma_{Residual0}$ = 60 ps in the case
where the retroreflector dispersion is 0.''

\begin{table}
\begin{center}
\begin{tabular}{|c||c|c|c|c|c|} \hline
8.0&  560&  670&  998& 1544& 2305\\ \hline
6.0&  341&  452&  787& 1342& 2116\\ \hline
4.0&  184&  297&  635& 1197& 1980\\ \hline
2.0&   89&  203&  544& 1110& 1899\\ \hline
0.0&   57&  171&  513& 1081& 1872\\ \hline \hline
 & 0.0& 2.0& 4.0& 6.0& 8.0\\ \hline
\end{tabular}
\end{center}
\caption{Minimum number of photons needed for a 1 mm measurement, as a function of libration angles, using existing equipment.  
Computed by taking the reflector dispersion from Table I, adding 50 ps in 
quadrature for the system response, then computing how many photons must be averaged to bring the rms value down to 6.6 ps (1 mm in range).}
\label{tab:photons}
\end{table}
The best measurements today have at least several mm of uncertainty.  
This is a product of two effects.  
First, each photon is uncertain because of retro-reflector dispersion.  
Second, there are few photons since the total link efficiency is so low.
To reduce the rms variation down to 6.6 ps (1 mm in range), 
in the straightforward way, thousands of returned photons must be averaged.
Table \ref{tab:photons} shows how many photons must be averaged to drive 
the rms accuracy down to 1 mm,
as a function of East-West and North-South librations.  
It assumes the A15 reflector and a 50 ps system response.
The Apache Point program\cite{Murphy1,Murphy2} hopes to accomplish the averaging of table \ref{tab:photons} directly, 
using a large telescope and good seeing as their primary tools.
\section{Taking advantage}
However, perhaps the reflector dispersion can be used to advantage.
First, note that by design a single corner cube will have the same optical 
path length no matter where on its surface the photon hits.  Even with the
inevitable manufacturing errors and thermal distortions, 
the response of a single corner cube will be nearly a delta function,  
just a tiny fraction of a ps wide.

\begin{figure*}
\begin{center}
\noindent
\includegraphics[height=7.0cm]{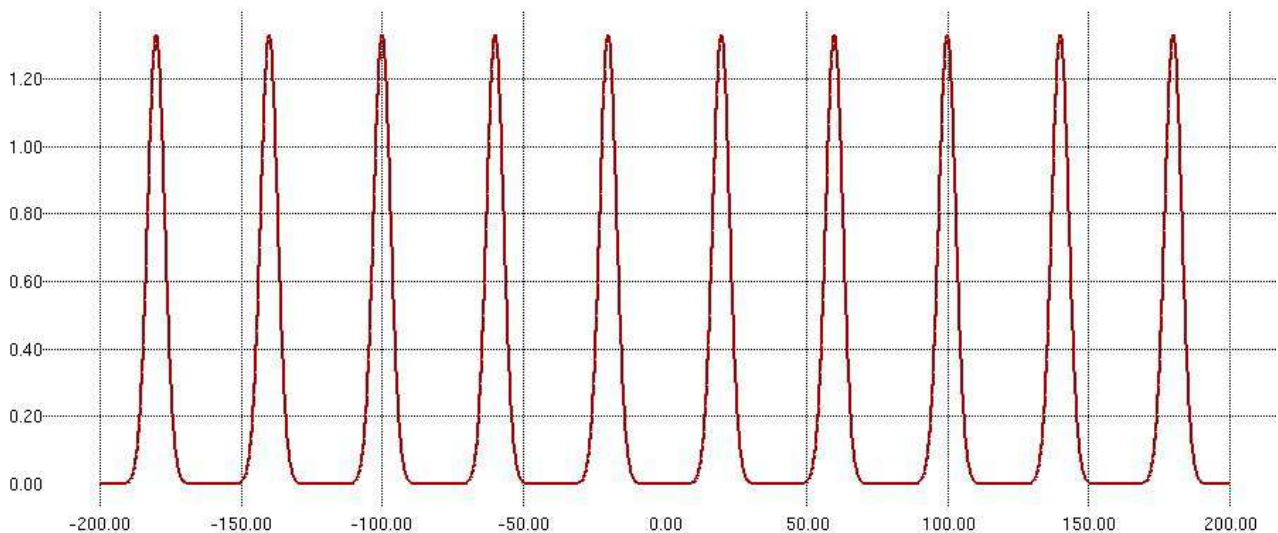}
\caption{Response of A11 (or A14) retro-reflector when East-West libration is 7.5 degrees, and North-South libration is 0, as measured from the reflector normal. Peaks have been made wider for clarity.}
\label{fig:TheoryResponse}
\end{center}
\end{figure*}

Next, as a thought experiment, imagine the case where the Earth, 
as seen by the A11 reflector array, is 7.5 degrees directly east or west.  
In this case the response of the detector depends only on which column of the array reflects each individual photon.  Each column is delayed from the next by
$$ \Delta t = D \cdot 2 \cdot \sin ( \alpha )/c$$
where $D$ is the distance between the columns, $\alpha$ is the angle off axis,  and $c$ the speed of light.  
For the Apollo 11 and 14 reflectors, $D$ is about 46 mm, and even if the reflector is perfectly aligned to the mean Earth position, 
the Earth will sometimes get at least $7.7^{\circ}$ off axis.  
This gives a time delay of 40 ps per column.  
Thus the overall response in this case consists of a comb comprised of 
10 delta functions, separated from each other by 40 ps.  Figure \ref{fig:TheoryResponse} shows this response, with the delta functions broadened for clarity.
Resolving individual retro-reflectors is not just a theoretical effect - it has been seen while ranging satellites.  
The Satellite Laser Ranging (SLR) station of the IWF, in Graz, Austria, has been able to resolve the individual retro-reflectors on several satellites, thanks to their short pulse (10 ps) KHz laser\cite{IWF2003,IWF2004,SingleRetro}.
From their 2004 annual report: `` 
Detecting single corner cubes of a retroreflector:
The kHz laser system now allows for
identifying single retro-reflectors, i.e., individual
corner cubes of the laser retroreflectors
for most of the satellites. This identification
can be done on the basis of the distribution
pattern of the return signals (Fig.
2.8), which is significantly different from the
pattern of special satellites, like e.g. the Russian
LaRetC, CHAMP, GRACE-A and GRACE-B,
which have been designed specifically with
only 1 retro being active at any time.''

\begin{figure*}
\begin{center}
\noindent
\includegraphics[height=7.5cm]{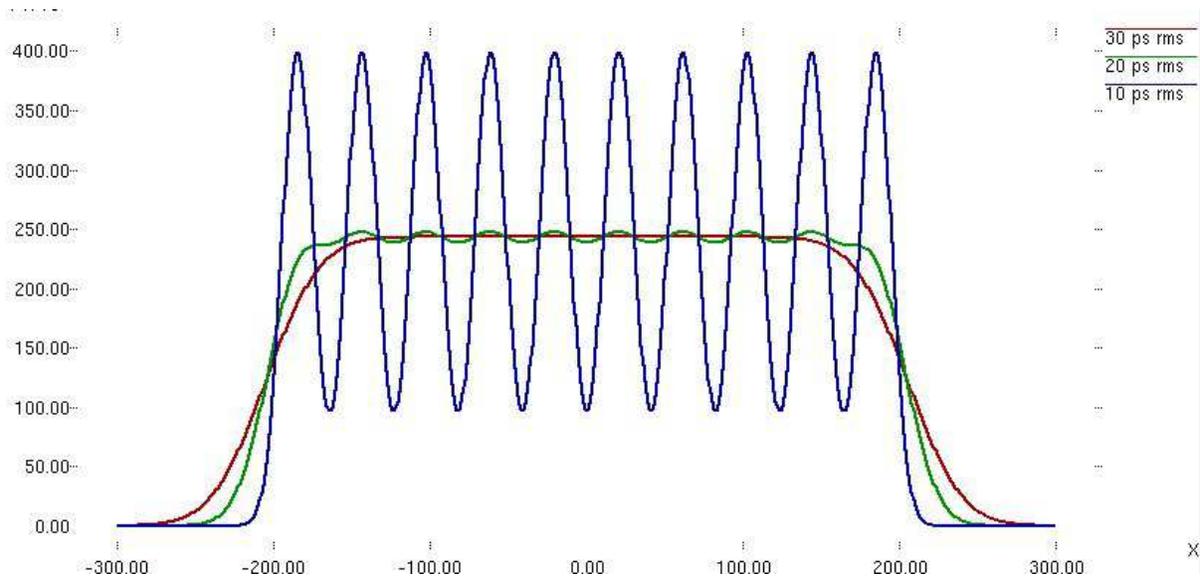}
\caption{Response of A11/A14 retro-reflector convolved with different system
resolutions}
\label{fig:response}
\end{center}
\end{figure*}

What would it take to resolve the corner cubes on the lunar retro-reflectors?  
The comb function response of the retroreflector array itself must be convolved
with the system response to determine the visibility of the fluctuations.
If the system response is 30 ps rms or greater, the comb function is 
completely washed out.  (See Figure \ref{fig:response}).  
At 20 ps rms, we begin to see wiggles in the response function, 
and at 10 ps the contrast is about 4:1 and can be easily resolved. 
\section{Previous work on resolving lunar reflectors}
There is a large body of work on lunar laser ranging\cite{over1,over2,over3}.
However, none of these appear to consider attempting to resolve the structure of
the reflectors.  
The Apache point effort\cite{Murphy1,Murphy2} explicitly relies on averaging
to get mm level measurements.  They note that their system should be able to
generate a range profile of each reflector, and measure, at least crudely, the
shape\cite{Murphy3} of the reflector.  
Resolving the reflector cubes is one more step in the same direction.

The OCA\cite{OCA} is also, indirectly measuring the reflector size (and shape, if
they measure the size at a number of libration conditions).  However, they also
treat reflector dispersion as an unavoidable obstacle:
``As long as almost all the echoes are obtained
only on the largest Apollo XV retroreflectors, a
major improvement of the LLR station precision is not
useful. ''

Another paper on possible improvements to Lunar Laser Ranging\cite{Tur04} concludes:
``The only way to improve measurement uncertainty in a passive array design is 
via a sparse array of small corner cubes.''  This seems unnecessarily pessimistic - the existing arrays are sparse enough.

The SLR community has long been interested in resolving the
individual corner cubes, and has done so in some cases\cite{IWF2003,IWF2004,SingleRetro}.

\section{Improving the response function}
What would it take to reduce the system response jitter to the point where the the modulation of the transfer function is visible?  
This seems hard but  possible  -
the satellite laser ranging community has been working towards mm precision for years.  
Much of this effort has gone into decreasing the system RMS (since they are not as photon poor as lunar ranging).
A study of atmosphere induced jitter by the SLR station in Graz, for example, measures an 8 ps RMS error to a fixed target 6 km away, and notes ``Close to the machine RMS'' \cite{Atmos1}.
This was done by (conceptually) straightforward improvements to the process - shorter pulses, a better event timer, and adjusting the pulse strength to minimize photodetector jitter.  Unfortunately the pulse strength that minimized the jitter was thousands of
photons per pulse, a pulse strength impractical for lunar ranging.

Can a very fast (10 ps or less) system response be obtained with any 
likely upgrade to the lunar laser facilities?
Looking at the error budgets of \cite{Murphy1,OCA}, we see three main impediments - 
the laser pulse width (typically 70-100 ps now), the detector response, (about 30-50 ps now), and the time-to-digital (TDC) converter, about 15 ps now.  

Clearly the first order of business is to reduce the width of the laser pulses.
Fortunately this is a well known technology.  
High peak power is very useful in many laser experiments, and lasers that 
seek high peak power do so by using very short pulse lengths (
a 1 joule pulse that is 1 ps long has a peak power of 1 terawatt). 
Therefore lasers similar to existing equipment, but with shorter pulses, can be bought off the shelf \cite{Laser1,Laser2}.  
These lasers have energies of 100mj or more per pulse, repetition rates of 
about 10 shots per second, and pulse widths of well under a ps.  

However, this straightforward approach will likely suffer from practical problems, 
such as destroying mirror coatings with the high power (terawatt peak) pulses.
Therefore it makes sense to move to KHz laser pulse rates, to keep peak power under 
control.  If we make the pulse 100 times narrower, but shoot 100 times more often with pulses with 100 times less energy per pulse, then the peak power and the number of returned photons should be unchanged.  
Also, many timing methods, including the one proposed in the next section, cannot take advantage of multiple photons per pulse.  
Therefore, especially with a large mirror and good seeing, 
KHz repetition rates using less powerful pulses would be helpful.

One concern is that with very short pulses, the exact wavelength becomes uncertain.  This seems manageable, since the shortest pulse that might be useful here (perhaps 1 ps) only has a spread in wavelength of about 0.3nm.  This is comparable to the widths of the narrowest filters used in existing systems and should not create a problem.

Some existing satellite ranging stations (such as Graz) have already moved to 
shorter pulses at higher repetition rates, so the problems are certainly solvable.

With a KHz pulse repetition rate, if the pulse repetition rate is fixed, sometimes the round trip delay will be exactly a multiple of the repetition rate.  
In this case the incoming pulse arrives just as the outgoing pulse departs, resulting in a host of technical difficulties.
However, this can always be avoided by changing the repetition rate slightly.  
For example, the round trip time will be 2.5 seconds exactly 
when the telescope and reflector are 374740.572 km apart.  This happens twice
a month as the moon passes from apogee to perigee and back.
At this distance, if the telescope launches pulses at exactly a 1 KHz rate, 
then the outgoing pulse will launch just as the return from 
the shot 2500 pulses ago comes in.  
Since the outgoing pulse is roughly $10^{17}$ times stronger, 
this is likely to cause interference.
The solution is to modify the pulse repetition rate slightly.
If we change the repetition rate to $2499.5/2.500 = 999.800$ Hz, then the round trip becomes a half integral number of pulses, and the returns are neatly sandwiched between the outgoing pulses.  
Changing the repetition rate is possible with existing short pulse technology since the pulses to be amplified are picked out of a train of much higher repetition rate by a high speed optical switch.

\subsection{Higher resolution detectors and timers}
The next problem is faster detectors.  This is a harder task.   
The fastest single photon detectors reported in the literature,
even in the laboratory, have about an 18 ps rms jitter.  
The timing precision can be improved by sacrificing sensitivity (such as the IWF work above) but these techniques are not useful for 
lunar ranging - the photons are too precious.
Timers are also a problem, but one that appears more easily solved - although the ones used for LLR are about 15ps rm jitter, better ones are known.  The Pico-Event Timer, used by the SLR community, has 1.2 ps resolution and 3 ps jitter\cite{PET}.

However, an alternative approach could get around both the diode response and timer problems simultaneously.  The basic idea is to time (rather than space) multiplex an array
of fast single photon photo-diodes.  If the optical beam dwells for 1 ps on each diode,
for example, then a simple determination of which diode was triggered measures the
incoming photon time to a ps.

What can be used to scan a light beam over an array of diodes at a sufficient rate and with high optical throughput?
A lithium niobate crystal in a time varying electric field can deflect a light beam from side to side at a quite high rate.  
For example, \cite{Deflect} shows an example of a linearized uni-directional deflector working at 15 GHz.  
Simple sine wave deflection at the same rate is correspondingly easier.  
Two crossed deflectors, with the appropriate phase shift, will move a light beam in a circle at the same 15 GHz rate.  
Then if we cover the circle with 64 photo-diodes, presumably SPAD devices, the beam 
dwells on each diode for 1 ps.  
When a photon comes in, it will trigger one of the 64 diodes.  
A conventional TDC, or cycle counting, can then identify the cycle, and which photo-diode tells which ps within the cycle.  
If the pulses are short, as proposed here, this technique cannot handle multiple photons per pulse - with 1 ps pulses, any multiple photon pulses would all hit the same SPAD,
and only the first would trigger it.
\begin{figure}
\begin{center}
\noindent
\includegraphics[height=6.5cm]{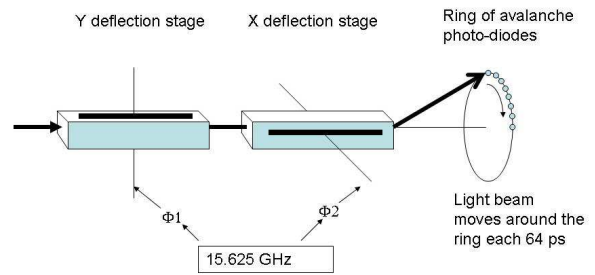}
\caption{Single photon detector with ps resolution.  This is a very simplified conceptual diagram - for example, the electrodes would need to be traveling wave structures to keep the E field in phase with the photon as it moves through the crystal. See \protect\cite{Deflect} for more realistic details.}
\label{fig:detector}
\end{center}
\end{figure}

This architecture, combined with short pulse widths, seems to lend itself to very accurate pulse travel time measurements.  
First, with a short enough pulse (where short enough means less than the dwell time on one photo diode), there is little problem with which photon within a pulse triggers the 
outgoing time measurement.  
With this scheme you could allow perhaps 20-30 photons from the outgoing pulse to hit the timer, so almost all outgoing pulses will have an accurate launch time.  
Next, the launch-return time is measured as an integral number of cycles, plus a time within the cycle depending on which detector responds.  
Both of these are unambiguous, down to the ps level.  
The remaining source of uncertainty is how well the microwave oscillator holds its frequency during the travel time of the pulse.
Modern microwave resonators\cite{Locke} have instability of $10^{-15}$ over a few second interval, and so would induce less than a ps of variance over the few second travel time.

This timing architecture has some other advantages not exploited here.
If used with longer pulses, this scheme can also count each of multiple photons per pulse, as long as there are fewer than about 8 per 64 ps.  This is similar to the use of SPAD arrays by Apollo, but using temporal rather than spatial multiplexing.  
If used on an astronomical source where photons arrive randomly, it could count them at up to 64 times the rate possible with a single SPAD, and with better time resolution.
\section{Simulations}
We assume throughout the experiments that the exact orientation of the
reflectors is known.  This is not currently the case, but if the corner cubes
can be resolved, it should be possible to determine the reflector orientations
by watching the transfer functions over a wide range of lunar libration angles.

Two strategies were simulated.  The first simplistic strategy just collects $N$ 
photons, and examines the fit.  
If the fit is good and unambiguous, a normal point is
generated, otherwise the data is discarded.  
This scheme would not be used in
practice, since it wastes photons, but it's easy to analyze.

The second strategy, an adaptive one, measures photons one at a time, 
until mm accuracy is achieved and there is only one plausible fit.
This is declared a normal point and then the next measurement starts.  
Each normal point may consist of a different number of photons.
If there is little or no structure to the return transfer function, 
this reduces to averaging.
But if we can resolve structure in the return, often many fewer photons will suffice.

Each measurement is simulated as follows.  
First, for each photon, we pick a corner cube.  
Then we pick a return time by adding in a random amount, chosen to match the distribution of the system response (here assumed to be gaussian).  
This gives an arrival time for each of the $N$ photons.  
Then we examine the goodness of fit to the ideal profile computed for that 
reflector at that time.  (Such an ideal profile is shown in 
figure \ref{fig:response}, for example).  

\section{Simulated results}
\begin{figure}[b]
\begin{center}
\noindent
\includegraphics[height=7.5cm]{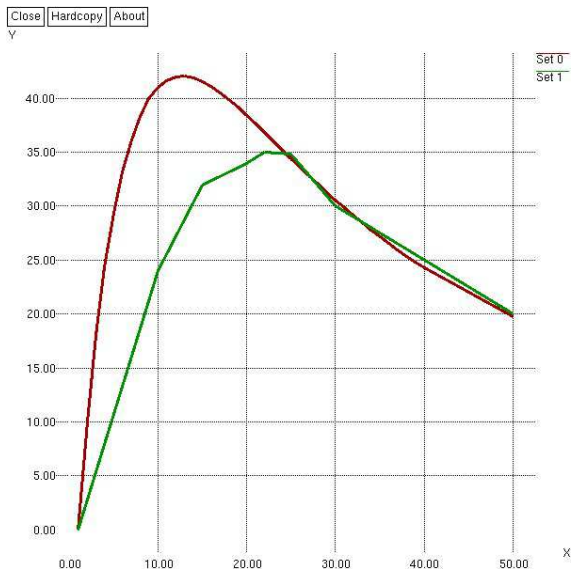}
\caption{Number of good measurements per 1000 photons (Y axis) as a function of number of photons in a measurement (X axis).  The upper curve is an optimistic guess requiring only at least one photon from the first and last columns.  The lower curve also requires a clearly unambiguous fit.  Note that above about 25 photons, the only significant source of ambiguous fits is missing first/last columns.}
\label{fig:GoodMeasure}
\end{center}
\end{figure}
First, look at the $N$ photon case.
We find two kinds of errors can creep into the fitting process.

In the first case, we can have a measurement that is very accurate, but ambiguous.
As an example, imagine the A11 array seen with the north-south angle equal to zero, 
and the east-west angle near its maximum.  
In this case only the column hit determines the delay.
In the fixed number of photon case, it is possible that none of the $N$ photons 
is reflected from either the first or last column (or both).  
In this case the fit is good but ambiguous - an (almost) equally good fit can be obtained by shifting the response one column.  
For this simple case, if $N$ photons hit $m$ columns at random, the probability $p$ that the first, or last, or both are empty is
$$ p = 2*\left(\frac{m-1}{m}\right)^N - \left(\frac{m-2}{m}\right)^N $$

In the second case we can simply be unlucky with our photon statistics, and 
no fit is particularly good.  This is quite common with small numbers of photons,
particularly if the attainable contrast is not high.  
We discard these points as well.

After some experimentation, we used the following heuristic - first, find the best fit, and measure its width to determine the precision.  
Then look for other plausible solutions.  If there is any other solutions, more than 2 standard deviations away and more than 20\% as good (measured as probability) then
we rejected that measurement as ambiguous.  
This test catches all the missing first/last column cases above, and others where the right solution is hard to discern because of the small number of data points.

The simulations were done with a 10 ps system response and the A11 reflector.
Above about 25 photons (with the parameters of the experiment) the odds of a plausible but wrong fit become negligible, and the bad solutions are dominated by the missing first/last column.  
As $N$, the number of photons per measurement,  increases we get fewer 
$N$ photon trials per 1000 photons, but each trial is less likely to suffer from ambiguity or the missing first/last column problem.
This tradeoff is shown in Fig. \ref{fig:GoodMeasure} where we show the number of 
normal points obtained per 1000 measured photons. (7.7 degrees east libration, 0 north/south libration, 10 ps system response, Apollo 11 reflector)
For this (very good) case the sweet spot is about 20-30 photons per measurement.

\subsection{Results with the adaptive algorithm}
The remaining experiments were done with the adaptive algorithm.  This is more likely to be used in practice since it does not waste precious photons.
The simulations were done with a 10 ps system response, the three different retro-reflector designs, and a range of misalignments.

There are several questions that naturally arise from this technique.  
How often, and by how much, does this technique really help?  
How often, during each lunar month, will we find the reflector at `good' positions?  To
attempt to answer these questions, we created a contour plot showing the gain compared to the existing techniques.  Since the reflectors are different, there is one plot per
reflector type (three in all).  
Superimposed on this plot are three `typical' libration trajectories, 
with marks a day apart, to help visualize how long the moon stays in a particular part of the plot.
Because the moon's perigee precesses with an 8.5 year period, the phase between the
North-South and East-West librations can assume any value from completely in phase to
completely out of phase.  Three libration trajectories are plotted on the diagram - 
in phase, out of phase ($90^{\circ}$ offset), and a $45^{\circ}$ offset.

\begin{figure*}[tb]
\begin{center}
\noindent
\includegraphics[height=12cm]{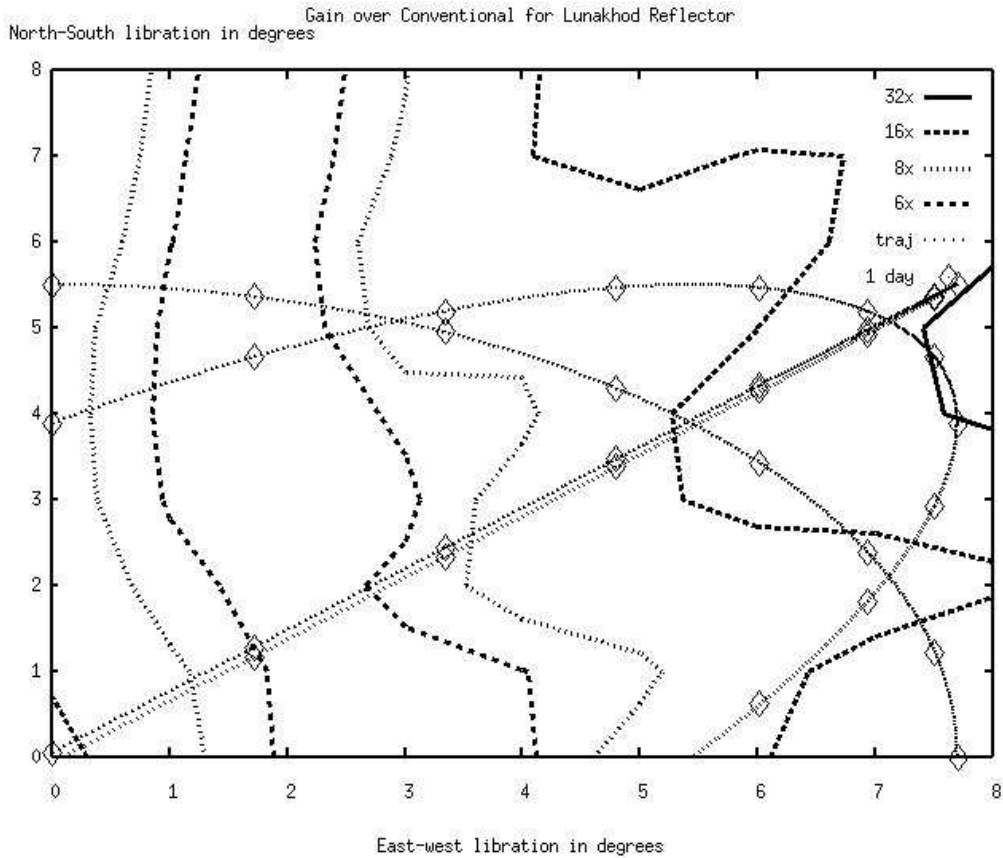}
\caption{Plot of gain as a function of libration angles for the Lunakhod reflector.  The contours show the gain (in photon number) over a conventional approach using averaging and a 50 ps system response.  The portions of ellipses are `typical' librations, shown where the apogee/perigee and North/South librations are in phase, out of phase, and halfway between.  The diamonds are one day apart on these curves.}
\label{fig:lrslt}
\end{center}
\end{figure*}

The biggest gains are observed with the Lunakhod reflector, as shown in table
\ref{tab:lrslt} and Fig. \ref{fig:lrslt}.
Fewer than 30 photons are needed for mm precision for any libration angles.  
At good angles, many fewer yet will do.  For comparison, raw averaging often takes 100 photons or
more, though short pulses alone could help this since the reflector is physically small.
\begin{table}[tbp]
\begin{center}
\setlength\fboxrule{0pt}
\small
\begin{tabular}{|c||c|c|c|c|c|c|c|c|c|} \hline
8&\fbox{\shortstack{7\\48\\103}}&\fbox{\shortstack{17\\51\\107}}&\fbox{\shortstack{28\\62\\117}}&\fbox{\shortstack{18\\80\\135}}&\fbox{\shortstack{11\\106\\161}}&\fbox{\shortstack{10\\138\\193}}&\fbox{\shortstack{10\\178\\233}}&\fbox{\shortstack{13\\224\\279}}&\fbox{\shortstack{15\\278\\333}} \\ \hline
7&\fbox{\shortstack{7\\37\\92}}&\fbox{\shortstack{16\\41\\96}}&\fbox{\shortstack{25\\52\\107}}&\fbox{\shortstack{15\\70\\125}}&\fbox{\shortstack{10\\96\\151}}&\fbox{\shortstack{10\\128\\183}}&\fbox{\shortstack{15\\168\\223}}&\fbox{\shortstack{17\\215\\270}}&\fbox{\shortstack{13\\269\\324}} \\ \hline
6&\fbox{\shortstack{8\\28\\83}}&\fbox{\shortstack{15\\32\\87}}&\fbox{\shortstack{21\\43\\98}}&\fbox{\shortstack{12\\61\\116}}&\fbox{\shortstack{11\\87\\142}}&\fbox{\shortstack{16\\120\\175}}&\fbox{\shortstack{20\\160\\215}}&\fbox{\shortstack{14\\207\\262}}&\fbox{\shortstack{11\\262\\317}} \\ \hline
5&\fbox{\shortstack{8\\20\\75}}&\fbox{\shortstack{15\\24\\79}}&\fbox{\shortstack{21\\35\\90}}&\fbox{\shortstack{12\\53\\109}}&\fbox{\shortstack{15\\79\\134}}&\fbox{\shortstack{20\\113\\168}}&\fbox{\shortstack{14\\153\\208}}&\fbox{\shortstack{10\\201\\256}}&\fbox{\shortstack{9\\256\\311}} \\ \hline
4&\fbox{\shortstack{8\\14\\69}}&\fbox{\shortstack{14\\17\\73}}&\fbox{\shortstack{21\\29\\84}}&\fbox{\shortstack{16\\47\\102}}&\fbox{\shortstack{19\\73\\128}}&\fbox{\shortstack{12\\107\\162}}&\fbox{\shortstack{10\\147\\202}}&\fbox{\shortstack{9\\195\\250}}&\fbox{\shortstack{9\\251\\306}} \\ \hline
3&\fbox{\shortstack{7\\9\\64}}&\fbox{\shortstack{12\\12\\68}}&\fbox{\shortstack{18\\24\\79}}&\fbox{\shortstack{18\\42\\97}}&\fbox{\shortstack{13\\68\\124}}&\fbox{\shortstack{11\\102\\157}}&\fbox{\shortstack{11\\143\\198}}&\fbox{\shortstack{12\\191\\246}}&\fbox{\shortstack{15\\247\\302}} \\ \hline
2&\fbox{\shortstack{6\\5\\60}}&\fbox{\shortstack{10\\9\\64}}&\fbox{\shortstack{16\\20\\75}}&\fbox{\shortstack{15\\39\\94}}&\fbox{\shortstack{13\\65\\120}}&\fbox{\shortstack{14\\99\\154}}&\fbox{\shortstack{21\\140\\195}}&\fbox{\shortstack{28\\188\\243}}&\fbox{\shortstack{22\\244\\299}} \\ \hline
1&\fbox{\shortstack{4\\4\\58}}&\fbox{\shortstack{8\\7\\62}}&\fbox{\shortstack{14\\18\\73}}&\fbox{\shortstack{18\\37\\92}}&\fbox{\shortstack{20\\63\\118}}&\fbox{\shortstack{22\\97\\152}}&\fbox{\shortstack{16\\138\\193}}&\fbox{\shortstack{12\\186\\241}}&\fbox{\shortstack{11\\242\\297}} \\ \hline
0&\fbox{\shortstack{3\\3\\57}}&\fbox{\shortstack{7\\7\\61}}&\fbox{\shortstack{13\\17\\72}}&\fbox{\shortstack{18\\36\\91}}&\fbox{\shortstack{22\\62\\117}}&\fbox{\shortstack{16\\96\\151}}&\fbox{\shortstack{13\\137\\192}}&\fbox{\shortstack{12\\186\\241}}&\fbox{\shortstack{11\\241\\297}} \\ \hline \hline
 & 0 & 1 & 2 & 3 & 4 & 5 & 6 & 7 & 8 \\ \hline
\end{tabular}
\caption{Lunakhod results.  Number of photons needed as a function of East-West libration in degrees (horizontal) and North-South libration (vertical).  Results are shown as A/B/C, where A is the number of photons needed (on the average) with short pulses and fitting.  B is the number of photons needed with short pulses and averaging, and C is the number of photons needed with system responses similar to existing facilities (about 50 ps).}
\label{tab:lrslt}
\end{center}
\end{table}

\begin{table}[tl]
\begin{center}
\small
\setlength\fboxrule{0pt}
\begin{tabular}{|c||c|c|c|c|c|c|c|c|c|} \hline
8&\fbox{\shortstack{16\\351\\406}}&\fbox{\shortstack{65\\356\\411}}&\fbox{\shortstack{91\\372\\427}}&\fbox{\shortstack{137\\398\\453}}&\fbox{\shortstack{201\\435\\490}}&\fbox{\shortstack{171\\482\\537}}&\fbox{\shortstack{124\\540\\595}}&\fbox{\shortstack{82\\608\\663}}&\fbox{\shortstack{74\\686\\741}} \\ \hline
7&\fbox{\shortstack{19\\270\\325}}&\fbox{\shortstack{63\\275\\330}}&\fbox{\shortstack{119\\291\\346}}&\fbox{\shortstack{146\\317\\372}}&\fbox{\shortstack{159\\354\\410}}&\fbox{\shortstack{168\\402\\457}}&\fbox{\shortstack{93\\460\\515}}&\fbox{\shortstack{74\\529\\584}}&\fbox{\shortstack{79\\608\\663}} \\ \hline
6&\fbox{\shortstack{22\\199\\254}}&\fbox{\shortstack{71\\204\\259}}&\fbox{\shortstack{110\\220\\275}}&\fbox{\shortstack{145\\247\\302}}&\fbox{\shortstack{155\\285\\340}}&\fbox{\shortstack{150\\333\\388}}&\fbox{\shortstack{81\\391\\446}}&\fbox{\shortstack{95\\460\\515}}&\fbox{\shortstack{114\\540\\595}} \\ \hline
5&\fbox{\shortstack{33\\139\\194}}&\fbox{\shortstack{65\\144\\199}}&\fbox{\shortstack{104\\161\\216}}&\fbox{\shortstack{130\\187\\243}}&\fbox{\shortstack{167\\225\\280}}&\fbox{\shortstack{97\\274\\329}}&\fbox{\shortstack{146\\333\\388}}&\fbox{\shortstack{151\\402\\457}}&\fbox{\shortstack{152\\482\\537}} \\ \hline
4&\fbox{\shortstack{33\\90\\145}}&\fbox{\shortstack{56\\95\\150}}&\fbox{\shortstack{85\\112\\167}}&\fbox{\shortstack{114\\139\\194}}&\fbox{\shortstack{161\\177\\232}}&\fbox{\shortstack{159\\225\\280}}&\fbox{\shortstack{153\\285\\340}}&\fbox{\shortstack{165\\354\\410}}&\fbox{\shortstack{215\\435\\490}} \\ \hline
3&\fbox{\shortstack{25\\52\\107}}&\fbox{\shortstack{43\\57\\112}}&\fbox{\shortstack{66\\73\\128}}&\fbox{\shortstack{86\\101\\156}}&\fbox{\shortstack{112\\139\\194}}&\fbox{\shortstack{133\\187\\243}}&\fbox{\shortstack{147\\247\\302}}&\fbox{\shortstack{151\\317\\372}}&\fbox{\shortstack{154\\398\\453}} \\ \hline
2&\fbox{\shortstack{18\\24\\79}}&\fbox{\shortstack{30\\30\\85}}&\fbox{\shortstack{45\\46\\101}}&\fbox{\shortstack{66\\73\\128}}&\fbox{\shortstack{84\\112\\167}}&\fbox{\shortstack{101\\161\\216}}&\fbox{\shortstack{111\\220\\275}}&\fbox{\shortstack{122\\291\\346}}&\fbox{\shortstack{97\\372\\427}} \\ \hline
1&\fbox{\shortstack{9\\9\\63}}&\fbox{\shortstack{16\\13\\68}}&\fbox{\shortstack{29\\30\\85}}&\fbox{\shortstack{43\\57\\112}}&\fbox{\shortstack{55\\95\\150}}&\fbox{\shortstack{65\\144\\199}}&\fbox{\shortstack{63\\204\\259}}&\fbox{\shortstack{56\\275\\330}}&\fbox{\shortstack{53\\356\\411}} \\ \hline
0&\fbox{\shortstack{3\\3\\57}}&\fbox{\shortstack{9\\9\\63}}&\fbox{\shortstack{16\\24\\79}}&\fbox{\shortstack{24\\52\\107}}&\fbox{\shortstack{31\\90\\145}}&\fbox{\shortstack{31\\139\\194}}&\fbox{\shortstack{21\\199\\254}}&\fbox{\shortstack{17\\270\\325}}&\fbox{\shortstack{15\\351\\406}} \\ \hline \hline
 & 0 & 1 & 2 & 3 & 4 & 5 & 6 & 7 & 8 \\ \hline
\end{tabular}
\caption{A11/A14 results.  Results are shown as A/B/C, where A is the number of photons needed (on the average) with short pulses and fitting.  B is the number of photons needed with short pulses and averaging, and C is the number of photons needed with system responses similar to existing facilities (about 50 ps).}
\label{tab:a11rslt}
\end{center}
\end{table}

\begin{figure*}[tb]
\begin{center}
\noindent
\includegraphics[height=12cm]{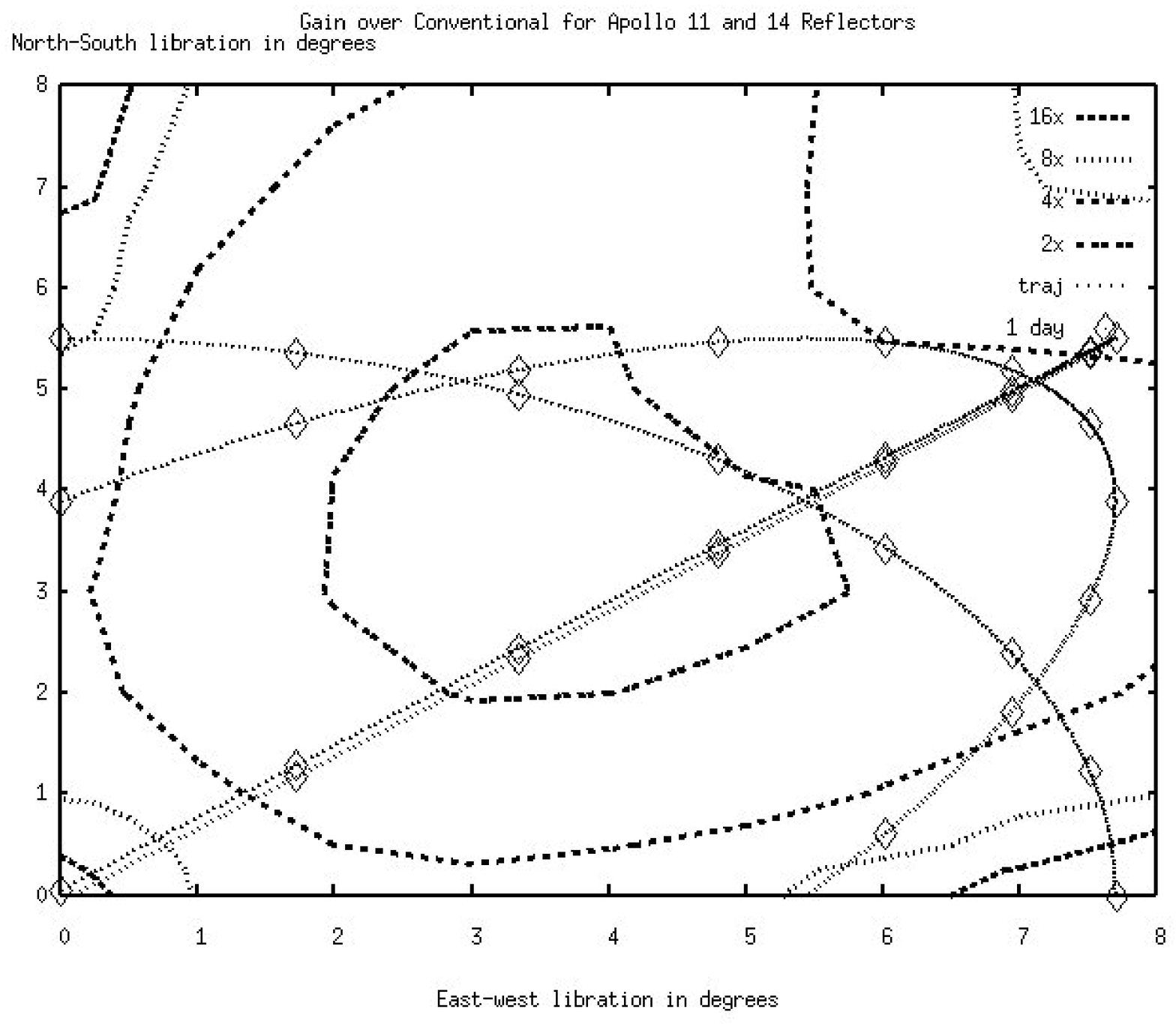}
\caption{Plot of gain as a function of libration angles for the A11/A14 reflectors.  The contours show the gain (in photon number) over a conventional approach using averaging and a 50 ps system response.  The portions of ellipses are `typical' librations, shown where the apogee/perigee and North/South librations are in phase, out of phase, and halfway between.  Diamonds are one day apart on these curves.}
\label{fig:a11rslt}
\end{center}
\end{figure*}

The next best gains are obtained with the A11 and A14 reflectors, as shown in
table \ref{tab:a11rslt} and Fig \ref{fig:a11rslt}.  
A factor of 2 over pure averaging is in general possible, with much higher 
gains when one of the two libration angles is near 0 (which generally happens 4 times per month).
Fewer than 150 photons are needed for mm precision for any libration angles.  At good angles, many fewer yet will do.

\begin{figure*}[b]
\begin{center}
\noindent
\includegraphics[height=12cm]{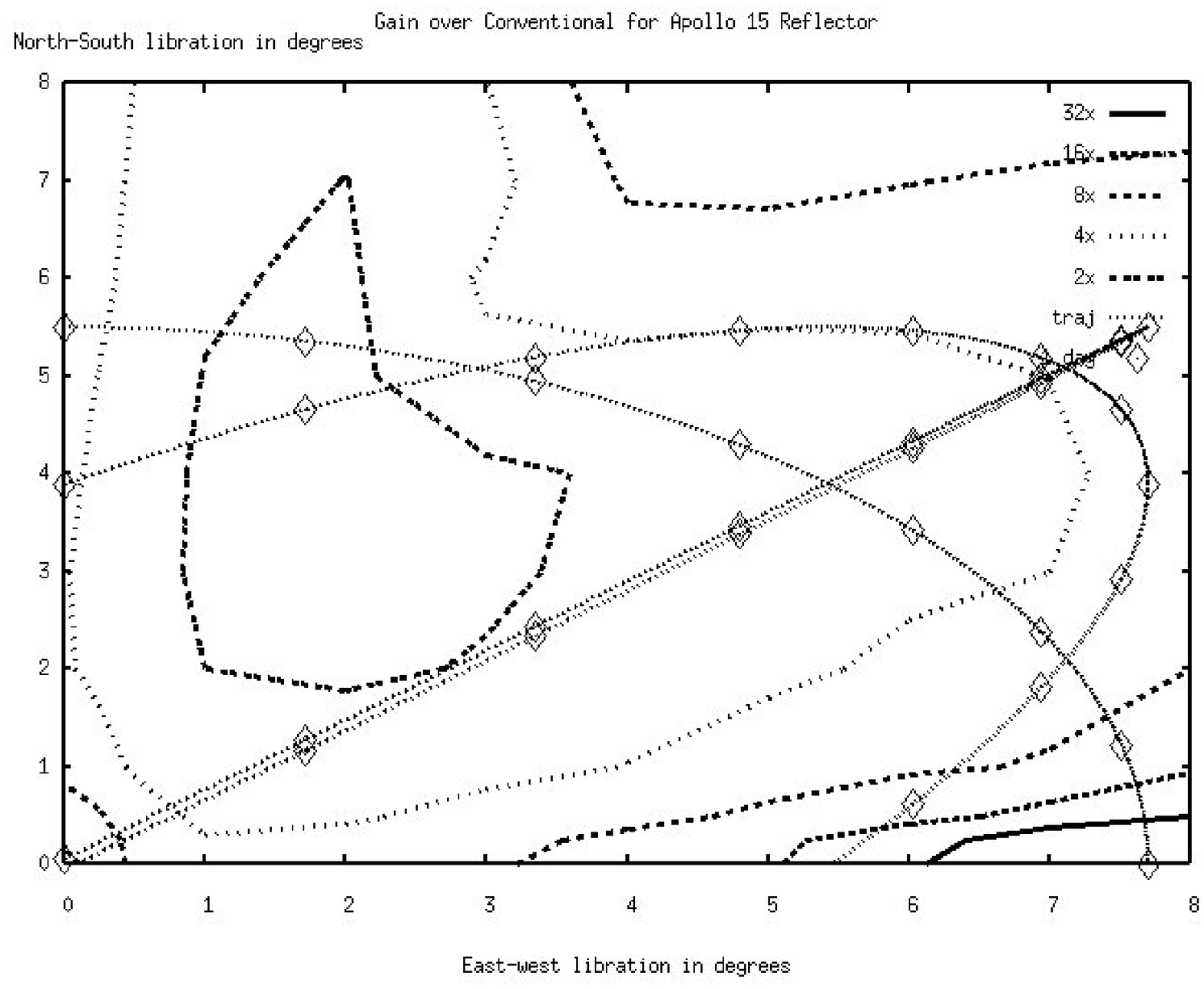}
\caption{Plot of gain as a function of libration angles, for the A15 reflector.  The contours show the gain (in photon number) over a conventional approach using averaging and a 50 ps system response.  The portions of ellipses are `typical' librations, shown where the apogee/perigee and North/South librations are in phase, out of phase, and halfway between.  The diamonds are one day apart on these curves.}
\label{fig:a15rslt}
\end{center}
\end{figure*}

Table \ref{tab:a15rslt}  and Fig \ref{fig:a15rslt} shows the improvement for the A15 reflector.  
This reflector shows the least improvement, due to its close packed layout, although 
fitting is still always better than averaging.  
The main exception is for very large East-West librations, 
where the improvement is a factor of 30 or so. 
This is because just one photon in the first and last column of the array serve to 
bracket the response and hence determine the centroid.  
The gap in the reflector also helps in this case, by providing an extra `first' and `last' column.
\begin{table}[tr]
\begin{center}
\small
\setlength\fboxrule{0pt}
\begin{tabular}{|c||c|c|c|c|c|c|c|c|c|} \hline
8&\fbox{\shortstack{77\\505\\560}}&\fbox{\shortstack{211\\532\\587}}&\fbox{\shortstack{286\\615\\670}}&\fbox{\shortstack{203\\752\\807}}&\fbox{\shortstack{94\\943\\998}}&\fbox{\shortstack{103\\1189\\1244}}&\fbox{\shortstack{135\\1489\\1544}}&\fbox{\shortstack{161\\1843\\1898}}&\fbox{\shortstack{220\\2250\\2305}} \\ \hline
7&\fbox{\shortstack{66\\388\\443}}&\fbox{\shortstack{189\\415\\470}}&\fbox{\shortstack{280\\498\\553}}&\fbox{\shortstack{241\\637\\692}}&\fbox{\shortstack{105\\830\\885}}&\fbox{\shortstack{124\\1078\\1133}}&\fbox{\shortstack{177\\1381\\1436}}&\fbox{\shortstack{251\\1738\\1793}}&\fbox{\shortstack{318\\2149\\2204}} \\ \hline
6&\fbox{\shortstack{59\\286\\341}}&\fbox{\shortstack{169\\314\\369}}&\fbox{\shortstack{266\\397\\452}}&\fbox{\shortstack{139\\537\\592}}&\fbox{\shortstack{127\\732\\787}}&\fbox{\shortstack{206\\982\\1037}}&\fbox{\shortstack{285\\1287\\1342}}&\fbox{\shortstack{335\\1647\\1702}}&\fbox{\shortstack{367\\2061\\2116}} \\ \hline
5&\fbox{\shortstack{50\\199\\255}}&\fbox{\shortstack{145\\227\\283}}&\fbox{\shortstack{232\\312\\367}}&\fbox{\shortstack{144\\452\\507}}&\fbox{\shortstack{268\\648\\703}}&\fbox{\shortstack{317\\900\\955}}&\fbox{\shortstack{369\\1207\\1262}}&\fbox{\shortstack{404\\1570\\1625}}&\fbox{\shortstack{363\\1986\\2042}} \\ \hline
4&\fbox{\shortstack{42\\129\\184}}&\fbox{\shortstack{118\\157\\212}}&\fbox{\shortstack{193\\242\\297}}&\fbox{\shortstack{272\\383\\438}}&\fbox{\shortstack{280\\580\\635}}&\fbox{\shortstack{342\\833\\888}}&\fbox{\shortstack{375\\1142\\1197}}&\fbox{\shortstack{432\\1506\\1561}}&\fbox{\shortstack{395\\1925\\1980}} \\ \hline
3&\fbox{\shortstack{32\\73\\128}}&\fbox{\shortstack{88\\102\\157}}&\fbox{\shortstack{153\\187\\242}}&\fbox{\shortstack{210\\329\\384}}&\fbox{\shortstack{257\\527\\582}}&\fbox{\shortstack{286\\781\\836}}&\fbox{\shortstack{327\\1091\\1146}}&\fbox{\shortstack{379\\1457\\1512}}&\fbox{\shortstack{360\\1878\\1933}} \\ \hline
2&\fbox{\shortstack{21\\34\\89}}&\fbox{\shortstack{59\\62\\117}}&\fbox{\shortstack{113\\148\\203}}&\fbox{\shortstack{166\\290\\345}}&\fbox{\shortstack{199\\489\\544}}&\fbox{\shortstack{233\\744\\799}}&\fbox{\shortstack{247\\1055\\1110}}&\fbox{\shortstack{284\\1422\\1477}}&\fbox{\shortstack{237\\1844\\1899}} \\ \hline
1&\fbox{\shortstack{11\\11\\65}}&\fbox{\shortstack{36\\39\\94}}&\fbox{\shortstack{68\\124\\179}}&\fbox{\shortstack{97\\267\\322}}&\fbox{\shortstack{129\\466\\521}}&\fbox{\shortstack{147\\721\\776}}&\fbox{\shortstack{157\\1033\\1088}}&\fbox{\shortstack{170\\1401\\1456}}&\fbox{\shortstack{133\\1823\\1879}} \\ \hline
0&\fbox{\shortstack{3\\3\\57}}&\fbox{\shortstack{20\\31\\86}}&\fbox{\shortstack{34\\116\\171}}&\fbox{\shortstack{42\\259\\314}}&\fbox{\shortstack{53\\458\\513}}&\fbox{\shortstack{54\\714\\769}}&\fbox{\shortstack{38\\1026\\1081}}&\fbox{\shortstack{27\\1394\\1449}}&\fbox{\shortstack{23\\1817\\1872}} \\ \hline \hline
 & 0 & 1 & 2 & 3 & 4 & 5 & 6 & 7 & 8 \\ \hline
\end{tabular}
\caption{Apollo 15 results.  Results are shown as A/B/C, where A is the number of photons needed (on the average) with short pulses and fitting.  B is the number of photons needed with short pulses and averaging, and C is the number of photons needed with system responses similar to existing facilities (about 50 ps).}
\label{tab:a15rslt}
\end{center}
\end{table}

\section{Experimental verification and future possibilities}
How can these techniques be tested?
Existing LLR stations may be able to image the largest single feature, 
the gap between the panels of the Apollo 15 retro-reflector, 
when seen at large East-West libration
and zero North-South libration.  This depends on the exact size of the gap,
which is not clear from any documentation the author could find.  

A more thorough conceptual test could be arranged relatively easily, 
by ranging a model of one of the lunar
retro-reflectors mounted several km from a state of the art SLR station (the SLR stations tend to have faster system responses than the LLR stations).  
Presumably it will be necessary to attenuate the return to the single photon per shot 
level, to avoid blocking the detector with the first returned photon.  
Under these conditions, existing SLR stations should be able to resolve the
individual corner cubes in the reflectors.
The station at Graz, for example, reports a 20 ps or better system rms at the
single photon level\cite{mm}.  
This is more than good enough to resolve the Lunakhod reflector
at realistic tilts, and the Apollo 11/14 reflectors could be resolved if larger tilts
are used for the sake of easy experimental testing.
Then given the measured timing for a number of photons, 
it would be easy to try the fitting procedure used here, and see if it works.  

Though much easier said than done, it appears the Graz station laser, detector, 
and timers, if coupled with a larger telescope, 
could routinely resolve the Lunakhod reflector.

If experiments prove positive, it might make sense to think about upgrading the the lunar ranging stations.
\section{Conclusions}
We have outlined a technique that may be able to give more accurate lunar ranges with fewer photons - dramatically fewer in some cases.  The necessary technical improvements seem to be within reach, although not easy.

This approach is complementary to traditional efforts.  It works best at large libration 
angles, where traditional averaging is weakest.  It works best on reflectors with the largest spacing, not those with the largest areas.  Finally, it relies on high tech lasers 
and timing to get mm class results, rather than large apertures and excellent seeing.

\bibliography{llr}

\begin{thebibliography}{20}
\expandafter\ifx\csname natexlab\endcsname\relax\def\natexlab#1{#1}\fi
\expandafter\ifx\csname bibnamefont\endcsname\relax
  \def\bibnamefont#1{#1}\fi
\expandafter\ifx\csname bibfnamefont\endcsname\relax
  \def\bibfnamefont#1{#1}\fi
\expandafter\ifx\csname citenamefont\endcsname\relax
  \def\citenamefont#1{#1}\fi
\expandafter\ifx\csname url\endcsname\relax
  \def\url#1{\texttt{#1}}\fi
\expandafter\ifx\csname urlprefix\endcsname\relax\def\urlprefix{URL }\fi
\providecommand{\bibinfo}[2]{#2}
\providecommand{\eprint}[2][]{\url{#2}}

\bibitem[{\citenamefont{Samain et~al.}(1998)\citenamefont{Samain, Mangin,
  Veillet, Torre, Fridelance, Chabaudie, F'eraudy, Glentzlin, Van, Furia
  et~al.}}]{OCA}
\bibinfo{author}{\bibfnamefont{E.}~\bibnamefont{Samain}},
  \bibinfo{author}{\bibfnamefont{J.}~\bibnamefont{Mangin}},
  \bibinfo{author}{\bibfnamefont{C.}~\bibnamefont{Veillet}},
  \bibinfo{author}{\bibfnamefont{J.}~\bibnamefont{Torre}},
  \bibinfo{author}{\bibfnamefont{P.}~\bibnamefont{Fridelance}},
  \bibinfo{author}{\bibfnamefont{J.}~\bibnamefont{Chabaudie}},
  \bibinfo{author}{\bibfnamefont{D.}~\bibnamefont{F'eraudy}},
  \bibinfo{author}{\bibfnamefont{M.}~\bibnamefont{Glentzlin}},
  \bibinfo{author}{\bibfnamefont{J.~P.} \bibnamefont{Van}},
  \bibinfo{author}{\bibfnamefont{M.}~\bibnamefont{Furia}},
  \bibnamefont{et~al.}, \bibinfo{journal}{Astronomy \& Astrophysics, supplement
  series} \textbf{\bibinfo{volume}{130}}, \bibinfo{pages}{235}
  (\bibinfo{year}{1998}).

\bibitem[{\citenamefont{Torre et~al.}(2004)\citenamefont{Torre, Furia, Mangin,
  and Samain}}]{Lunokhod1}
\bibinfo{author}{\bibfnamefont{J.}~\bibnamefont{Torre}},
  \bibinfo{author}{\bibfnamefont{M.}~\bibnamefont{Furia}},
  \bibinfo{author}{\bibfnamefont{J.}~\bibnamefont{Mangin}}, \bibnamefont{and}
  \bibinfo{author}{\bibfnamefont{E.}~\bibnamefont{Samain}}, in
  \emph{\bibinfo{booktitle}{Proceedings of the 14th International Laser Ranging
  Workshop, San Fernando, Spain}} (\bibinfo{year}{2004}).

\bibitem[{MyS()}]{MySpace}
\emph{\bibinfo{title}{{Apollo Lunar Surface Experiment Package (ALSEP)}}},
  \urlprefix\url{http://www.myspacemusuem.com/alseph1.htm}.

\bibitem[{\citenamefont{T.~W.~Murphy
  et~al.}({\natexlab{a}})\citenamefont{T.~W.~Murphy, Strasburg, Stubbs,
  Adelberger, and Angle}}]{Murphy1}
\bibinfo{author}{\bibfnamefont{J.}~\bibnamefont{T.~W.~Murphy}},
  \bibinfo{author}{\bibfnamefont{J.~D.} \bibnamefont{Strasburg}},
  \bibinfo{author}{\bibfnamefont{C.~W.} \bibnamefont{Stubbs}},
  \bibinfo{author}{\bibfnamefont{E.~G.} \bibnamefont{Adelberger}},
  \bibnamefont{and} \bibinfo{author}{\bibfnamefont{J.}~\bibnamefont{Angle}},
  \emph{\bibinfo{title}{The {A}pache point observatory lunar laser-ranging
  operation ({APOLLO})}},
  \urlprefix\url{physics.ucsd.edu/~tmurphy/apollo/doc/matera.pdf}.

\bibitem[{\citenamefont{T.~W.~Murphy
  et~al.}({\natexlab{b}})\citenamefont{T.~W.~Murphy, Adelberger, Strasburg, and
  Stubbs}}]{Murphy2}
\bibinfo{author}{\bibfnamefont{J.}~\bibnamefont{T.~W.~Murphy}},
  \bibinfo{author}{\bibfnamefont{E.~G.} \bibnamefont{Adelberger}},
  \bibinfo{author}{\bibfnamefont{J.~D.} \bibnamefont{Strasburg}},
  \bibnamefont{and} \bibinfo{author}{\bibfnamefont{C.~W.}
  \bibnamefont{Stubbs}}, \emph{\bibinfo{title}{{APOLLO}: Multiplexed lunar
  laser ranging}},
  \urlprefix\url{http://physics.ucsd.edu/~tmurphy/apollo/doc/multiplex.pdf}.

\bibitem[{IWF(2003)}]{IWF2003}
\emph{\bibinfo{title}{Annual report 2003, space research institute {Graz},
  austrian academy of sciences}} (\bibinfo{year}{2003}),
  \urlprefix\url{http://www.iwf.oeaw.ac.at/files/iwf_ar_2003.pdf}.

\bibitem[{IWF(2004)}]{IWF2004}
\emph{\bibinfo{title}{Annual report 2004, space research institute {Graz},
  austrian academy of sciences}} (\bibinfo{year}{2004}),
  \urlprefix\url{http://www.iwf.oeaw.ac.at/files/iwf_ar_2004.pdf}.

\bibitem[{\citenamefont{Arnold et~al.}(2004)\citenamefont{Arnold, Kirchner, and
  Koidl}}]{SingleRetro}
\bibinfo{author}{\bibfnamefont{D.}~\bibnamefont{Arnold}},
  \bibinfo{author}{\bibfnamefont{G.}~\bibnamefont{Kirchner}}, \bibnamefont{and}
  \bibinfo{author}{\bibfnamefont{F.}~\bibnamefont{Koidl}}, in
  \emph{\bibinfo{booktitle}{Proceedings of the 14th International Laser Ranging
  Workshop, San Fernando, Spain}} (\bibinfo{year}{2004}).

\bibitem[{\citenamefont{Bender et~al.}(1973)\citenamefont{Bender, Currie,
  Dicke, Eckhardt, Faller, Kaula, Mulholland, Plotkin, Poultney, Silverberg
  et~al.}}]{over1}
\bibinfo{author}{\bibfnamefont{P.~L.} \bibnamefont{Bender}},
  \bibinfo{author}{\bibfnamefont{D.~G.} \bibnamefont{Currie}},
  \bibinfo{author}{\bibfnamefont{R.~H.} \bibnamefont{Dicke}},
  \bibinfo{author}{\bibfnamefont{D.~H.} \bibnamefont{Eckhardt}},
  \bibinfo{author}{\bibfnamefont{J.~E.} \bibnamefont{Faller}},
  \bibinfo{author}{\bibfnamefont{W.~M.} \bibnamefont{Kaula}},
  \bibinfo{author}{\bibfnamefont{J.~D.} \bibnamefont{Mulholland}},
  \bibinfo{author}{\bibfnamefont{H.~H.} \bibnamefont{Plotkin}},
  \bibinfo{author}{\bibfnamefont{S.~K.} \bibnamefont{Poultney}},
  \bibinfo{author}{\bibfnamefont{E.~C.} \bibnamefont{Silverberg}},
  \bibnamefont{et~al.}, \bibinfo{journal}{Science}
  \textbf{\bibinfo{volume}{182}}, \bibinfo{pages}{229} (\bibinfo{year}{1973}).

\bibitem[{\citenamefont{Dickey et~al.}(1994)\citenamefont{Dickey, Bender,
  Faller, Newhall, Ricklefs, Ries, Shelus, Veillet, Whipple, Wiant
  et~al.}}]{over2}
\bibinfo{author}{\bibfnamefont{J.~O.} \bibnamefont{Dickey}},
  \bibinfo{author}{\bibfnamefont{P.~L.} \bibnamefont{Bender}},
  \bibinfo{author}{\bibfnamefont{J.~E.} \bibnamefont{Faller}},
  \bibinfo{author}{\bibfnamefont{X.~X.} \bibnamefont{Newhall}},
  \bibinfo{author}{\bibfnamefont{R.~L.} \bibnamefont{Ricklefs}},
  \bibinfo{author}{\bibfnamefont{J.~G.} \bibnamefont{Ries}},
  \bibinfo{author}{\bibfnamefont{P.~J.} \bibnamefont{Shelus}},
  \bibinfo{author}{\bibfnamefont{C.}~\bibnamefont{Veillet}},
  \bibinfo{author}{\bibfnamefont{A.~L.} \bibnamefont{Whipple}},
  \bibinfo{author}{\bibfnamefont{J.~R.} \bibnamefont{Wiant}},
  \bibnamefont{et~al.}, \bibinfo{journal}{Science} pp.
  \bibinfo{pages}{482--490} (\bibinfo{year}{1994}).

\bibitem[{ove()}]{over3}
\emph{\bibinfo{title}{History of laser ranging and {MLRS}}},
  \urlprefix\url{http://www.csr.utexas.edu/mlrs/history.html}.

\bibitem[{\citenamefont{Murphy}(2002)}]{Murphy3}
\bibinfo{author}{\bibfnamefont{T.}~\bibnamefont{Murphy}},
  \emph{\bibinfo{title}{Apollo: Multiplexed lunar laser ranging}}
  (\bibinfo{year}{2002}),
  \urlprefix\url{http://cddis.gsfc.nasa.gov/lw13/docs/presentations/llr_murphy%
_1p.pdf}.

\bibitem[{\citenamefont{Turyshev et~al.}(2004)\citenamefont{Turyshev, Williams,
  Shao, Anderson, Jr, and Jr.}}]{Tur04}
\bibinfo{author}{\bibfnamefont{S.~G.} \bibnamefont{Turyshev}},
  \bibinfo{author}{\bibfnamefont{J.~G.} \bibnamefont{Williams}},
  \bibinfo{author}{\bibfnamefont{M.}~\bibnamefont{Shao}},
  \bibinfo{author}{\bibfnamefont{J.~D.} \bibnamefont{Anderson}},
  \bibinfo{author}{\bibfnamefont{K.~L.~N.} \bibnamefont{Jr}}, \bibnamefont{and}
  \bibinfo{author}{\bibfnamefont{T.~W.~M.} \bibnamefont{Jr.}},
  \emph{\bibinfo{title}{Laser ranging to the moon, mars and beyond}}
  (\bibinfo{year}{2004}), \urlprefix\url{http://arxiv.org/gr-qc/0411082}.

\bibitem[{\citenamefont{Mulacova et~al.}(2004)\citenamefont{Mulacova, Hamal,
  Kirchner, and Koidl}}]{Atmos1}
\bibinfo{author}{\bibfnamefont{J.}~\bibnamefont{Mulacova}},
  \bibinfo{author}{\bibfnamefont{K.}~\bibnamefont{Hamal}},
  \bibinfo{author}{\bibfnamefont{G.}~\bibnamefont{Kirchner}}, \bibnamefont{and}
  \bibinfo{author}{\bibfnamefont{F.}~\bibnamefont{Koidl}}, in
  \emph{\bibinfo{booktitle}{Proceedings of the 14th International Laser Ranging
  Workshop, San Fernando, Spain}} (\bibinfo{year}{2004}),
  \bibinfo{note}{http://cddis.gsfc.nasa.gov/lw14/docs/presnts/atm3\_khp.pdf}.

\bibitem[{Las({\natexlab{a}})}]{Laser1}
\urlprefix\url{http://www.cohr.com/Downloads/Custom_Laser_Brochure.pdf}.

\bibitem[{Las({\natexlab{b}})}]{Laser2}
\urlprefix\url{http://www.femtosecondsystems.com/products/femtosecond_lasers/c%
r_forsterite_lasers/jaws/jaws.php/129}.

\bibitem[{\citenamefont{K.Hamal and I.Prochazka}(1998)}]{PET}
\bibinfo{author}{\bibnamefont{K.Hamal}} \bibnamefont{and}
  \bibinfo{author}{\bibnamefont{I.Prochazka}}, \bibinfo{journal}{Annales
  Geophysicae Supplement} \textbf{\bibinfo{volume}{16}} (\bibinfo{year}{1998}).

\bibitem[{\citenamefont{Khayim et~al.}(2001)\citenamefont{Khayim, Maruko,
  Shibuya, Morimoto, and Kobayashi}}]{Deflect}
\bibinfo{author}{\bibfnamefont{T.}~\bibnamefont{Khayim}},
  \bibinfo{author}{\bibfnamefont{A.}~\bibnamefont{Maruko}},
  \bibinfo{author}{\bibfnamefont{K.}~\bibnamefont{Shibuya}},
  \bibinfo{author}{\bibfnamefont{A.}~\bibnamefont{Morimoto}}, \bibnamefont{and}
  \bibinfo{author}{\bibfnamefont{T.}~\bibnamefont{Kobayashi}},
  \bibinfo{journal}{IEEE Journal of Quantum Electronics}
  \textbf{\bibinfo{volume}{37}}, \bibinfo{pages}{964 } (\bibinfo{year}{2001}).

\bibitem[{\citenamefont{Locke et~al.}(2003)\citenamefont{Locke, Munro, Tobar,
  Ivanov, and Santarelli}}]{Locke}
\bibinfo{author}{\bibfnamefont{C.}~\bibnamefont{Locke}},
  \bibinfo{author}{\bibfnamefont{S.}~\bibnamefont{Munro}},
  \bibinfo{author}{\bibfnamefont{M.}~\bibnamefont{Tobar}},
  \bibinfo{author}{\bibfnamefont{E.}~\bibnamefont{Ivanov}}, \bibnamefont{and}
  \bibinfo{author}{\bibfnamefont{G.}~\bibnamefont{Santarelli}}, in
  \emph{\bibinfo{booktitle}{Proceedings of the 2003 IEEE International
  Frequency Control Symposium and PDA Exhibition}} (\bibinfo{year}{2003}).

\bibitem[{\citenamefont{Kral et~al.}(2004)\citenamefont{Kral, Hamal, Prochazka,
  Kirchner, and Koidl}}]{mm}
\bibinfo{author}{\bibfnamefont{L.}~\bibnamefont{Kral}},
  \bibinfo{author}{\bibfnamefont{K.}~\bibnamefont{Hamal}},
  \bibinfo{author}{\bibfnamefont{I.}~\bibnamefont{Prochazka}},
  \bibinfo{author}{\bibfnamefont{G.}~\bibnamefont{Kirchner}}, \bibnamefont{and}
  \bibinfo{author}{\bibfnamefont{F.}~\bibnamefont{Koidl}},
  \emph{\bibinfo{title}{Ground target khz laser ranging with submillimeter
  precision}} (\bibinfo{year}{2004}),
  \urlprefix\url{http://kral.astronomy.cz/doc/submmLR_Graz.ppt}.

\end{thebibliography}

\end{document}